\documentclass[multphys,vecphys]{svmult}

\usepackage{amsmath,amssymb,amsfonts,mathrsfs,latexsym} 
\usepackage{graphicx}
\usepackage{makeidx}
\usepackage{multicol}
\makeindex

\newcommand{\TT}{\mathrm{TT}}
\newcommand{\B}{\mathrm{B}}
\newcommand{\GW}{\mathrm{GW}}
\newcommand{\Ibar}{{\cal I} \kern -0.5em\raise 0.25ex \hbox{--}}

\newcommand{\be}{\begin{equation}}
\newcommand{\ee}{\end{equation}}

\begin{document}

\title*{Gravitational-wave astronomy: the high-frequency window}

\author{Nils Andersson\inst{1} \and Kostas D Kokkotas \inst{2}}

\institute{ School of Mathematics, University of Southampton,
Southampton SO17 1BJ, UK \texttt{N.Andersson@maths.soton.ac.uk}
\and Department of Physics, Aristotle University of Thessaloniki,
54124 Greece \& \\ Center for Gravitational Wave Physics, 
104 Davey Laboratory, University Park, PA 16802, USA. 
\texttt{kokkotas@auth.gr}}

\maketitle
\begin{abstract}
As several large scale interferometers are beginning to take 
data at sensitivities where astrophysical sources are predicted, 
the direct detection of gravitational waves may well be imminent. 
This would (finally) open the long anticipated gravitational-wave
window to our Universe, and should lead to a much improved
understanding of the most violent processes imaginable; 
the formation of black holes and neutron stars following core
collapse supernovae and the merger of compact objects at the end of 
binary inspiral. Over the next decade we can hope to learn much 
about the extreme physics associated with, in particular, neutron stars. 

This contribution is divided in two parts. The first part 
provides a text-book level introduction to 
gravitational radiation. The key concepts required for a discussion 
of gravitational-wave physics are introduced. In particular, the 
quadrupole formula is applied to the anticipated ``bread-and-butter'' source
for detectors like LIGO, GEO600, EGO and TAMA300: inspiralling
compact binaries. The second part    
provides a brief review of high frequency gravitational waves. 
In the frequency range above (say) 100~Hz, 
gravitational collapse, rotational instabilities and oscillations
of the remnant compact objects are potentially important sources
of gravitational waves. Significant and unique information concerning the
various stages of collapse, the evolution of protoneutron
stars and the details of the supranuclear equation of state of such objects
can be drawn from careful study of the gravitational-wave signal.
As the amount of exciting physics one may be able to study via 
the detections of gravitational waves from these sources is truly
inspiring, there is strong motivation for the development of future
generations of ground based detectors sensitive in the 
range from hundreds of Hz to several kHz.
\end{abstract}

\section{Introduction}

One of the central predictions of Einsteins' general theory of relativity 
is that gravitational waves will be generated as masses are
accelerated. Despite decades of effort these ripples in spacetime have still not 
been observed directly. Yet we have strong indirect evidence for their 
existence from the excellent agreement between the observed 
inspiral rate of the binary pulsar PSR1913+16 and the theoretical
prediction (better than 1\% 
in the phase evolution). 
This provides confidence in the theory and suggests that 
``gravitational-wave astronomy'' should be viewed as a
serious proposition. Provided that
i) detectors with the required sensitivity can be constructed, and  
ii) the significant data analysis challenge can be dealt with, 
this new window to the Universe promises to 
bring unprecedented insights into the most 
violent events imaginable; supernova explosions, binary mergers
and the big bang itself. A key reason for this
expectation follows from a comparison between 
gravitational and electromagnetic waves:

\begin{itemize}
\item While electromagnetic waves are radiated by individual particles, 
gravitational waves are due to non-spherical bulk motion of matter. 
In essence, this means that the information carried by electromagnetic 
waves is stochastic in nature, while the gravitational waves provide 
insights into coherent mass currents.   

\item The electromagnetic waves will have been 
scattered many times. In contrast, gravitational
waves interact weakly with matter
and arrive at the Earth in pristine condition. 
This means that gravitational waves can be used to probe regions
of space that are opaque to electromagnetic waves, It is, of course, 
a blessing in disguise since the weak interaction with matter also
makes the gravitational waves fiendishly hard to detect.

\item
 Standard astronomy is based on deep
imaging of small fields of view, while gravitational-wave 
detectors cover virtually the entire sky.  

\item Electromagnetic radiation  has a wavelength smaller
than the size of the emitter, while  the
wavelength of a gravitational wave is usually 
larger than the size of the source. This means that we cannot 
use gravitational-wave data to create an image of the source. 
In fact, gravitational-wave observations are more like 
audio than visual. 
\end{itemize}

\noindent
\emph{Morale:} Gravitational waves carry information
which would be very difficult to glean 
by other means.
By analysing gravitational-wave data we can expect to 
learn a lot about the extreme physics governing compact objects.
This should lead to answers to many outstanding questions in astrophysics, 
for example, 
\begin{itemize}

\item What is the black-hole population of the Universe? Observations of 
the nonlinear spacetime dynamics associated with binary merger, 
as well as the quasinormal mode (QNM) ringing which is likely to dominate the 
radiation at late times, should provide direct proof of the presence
of a black hole, as well as a measure of it's mass and spin. 

\item Are astrophysical black holes, indeed, described by the Kerr metric?
By studying the inspiral of a low-mass object into a supramassive
black hole (as anticipated at the centre of most galaxies) we can hope to 
construct a detailed map of the exterior black-hole spacetime.

\item What is the supranuclear neutron star equation of state?
This is a very difficult question, which potentially involves 
an understanding of the role of exotic phases of matter, like 
deconfined quarks, superfluidity/conductivity, extreme magnetic fields
etcetera. At an immediate level, gravitational-wave observations of 
the various oscillation modes of a compact star may be used to 
solve the ``inverse problem'' for parameters like the mass and the radius 
of the star. At a more subtle level, the spectrum of the star depends on 
the internal physics. For example, strong composition gradients
which affect the gravity g-modes significantly, and superfluidity 
which leads to the existence of new classes of oscillations associated 
with the more or less decoupled additional degrees of freedom
of such a system. Finally, the internal physics also plays a crucial
role in determining the extent to which various rotational
instabilities will be able to grow to an interesting amplitude. 
In the case of the Coriolis driven r-modes, their instability 
may be severely suppressed by the presence of hyperons in the stars
core. A key role is also played by the internal magnetic fields.  
\end{itemize} 

At the time of writing, the new generation of 
interferometric gravitational-wave detectors (in particular LIGO and GEO600)
is already collecting data at a  sensitivity at least one order
of magnitude better than that of the operating resonant detectors.
In the first instance, the broadband detectors will be
sensitive in a range of frequencies between 50 and
a few hundred Hz. This frequency window is of great interest 
since an inspiraling compact binary will move through it during the 
last few minutes before merger. Such sources are the natural 
``bread and butter'' source for the  detectors. 
The next generation of interferometers will broaden the bandwidth somewhat
but will still not be very sensitive to frequencies above
500-600~Hz, unless they are operated in a  narrow-band configuration
\cite{ThorneCutler,GEO600}. There are also interesting suggestions for wide
band resonant detectors in the kHz band \cite{Cerdonio}.
The move towards higher frequencies is 
driven by the wealth of exciting sources that radiate
in the range from a few hundred Hz up to several kHz.  

Our contribution to these proceedings is divided in 
two parts. The first part describes
gravitational-wave physics at an introductory level. The second part
provides a brief review of the main sources that radiate
in the frequency band above a few hundred Hz.
We believe that these sources are the natural targets for a
\underline{third} generation of ground based detectors. As we will discuss, 
there are a variety of sources associated with very
interesting physics in this high-frequency window. 
These sources
clearly deserve special attention, and if either resonant or narrow-band
interferometers can achieve the required sensitivity, a
plethora of unique information  can be gathered.

\section{Einstein's elusive waves}

The aim of the first part of our contribution is to provide a 
condensed text-book level introduction to  gravitational waves.
Although in no sense complete this description should prepare the 
reader for the discussion of high-frequency sources which follows. 

\subsection{The nature of the waves}

The first aspect of gravitational waves that we need to appreciate is their
\emph{tidal} nature. This is important because it implies that 
they can only be measured through the 
relative motion of bodies. That this should be the case is easy to 
understand. In general relativity we can always construct a local 
inertial frame associated with a given observer. In this local frame, 
spacetime will by construction be flat which means that we cannot 
hope to observe the local deformations which would correspond
to a gravitational wave.  
\begin{figure}
 \centerline{\includegraphics[width=5cm]{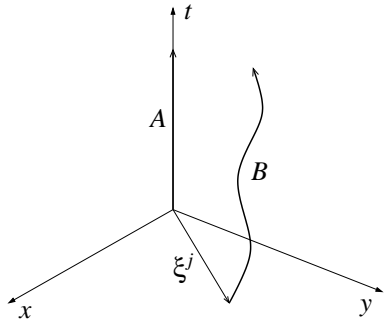}}
\caption{An illustration of the two test particles discussed in the text. 
If a gravitational wave passes through, particle $A$ will observe 
particle $B$ oscillating back and forth.
}
\label{fig1}
\end{figure}
Consider two test particles, $A$ and $B$, 
that are initially at rest. Assume that 
they are separated by a 
purely spatial vector
$\xi^j$ (hereafter latin indices are spatial and run from 1--3, while greek indices are spacetime and run from 0--4), 
and use the local inertial frame in which particle $A$ remains 
at the origin for the calculation.
In this case the equation of geodesic deviation can be written
\be
\frac{ \partial^2 \xi^j}{\partial t^2} = - R^j_{\ 0k0}\xi^k
\ee
where $R_{\mu \nu \delta \gamma}$ is the Riemann tensor. Here it represents the 
curvature induced by the gravitational wave. 
Letting $\xi^j = x_{0}^j + \delta x^j$, with $\delta x^j$ the small displacement away 
from the original position, we get
\be 
\frac{ \partial^2 \delta x^j}{\partial t^2} \approx - R^j_{\ 0k0} x_{0}^k
=  - R_{j0k0} x_{0}^k
\ee 
Now it is natural to define the {\em gravitational wave-field} $h_{jk}$ through
\be
R_{j0k0} \equiv - { 1\over2} {\partial^2 h_{jk} \over \partial t^2}
\longrightarrow
{ \partial^2 \delta x_j \over \partial t^2} \approx 
{ 1\over2} {\partial^2 h_{jk} \over \partial t^2} x_{0}^k
\ee
which integrates to 
\be
\delta x_j = {1\over 2} h_{jk}  x_{0}^k \quad \mbox{or} \quad
h \approx { \Delta L \over L}
\ee
where $h$ is the dimensionless gravitational-wave strain.
From this exercise we learn that, 
in order to detect gravitational waves we need to monitor (with extreme
precision) the relative motion 
of test masses.
\begin{figure}
 \centerline{\includegraphics[width=8.5cm]{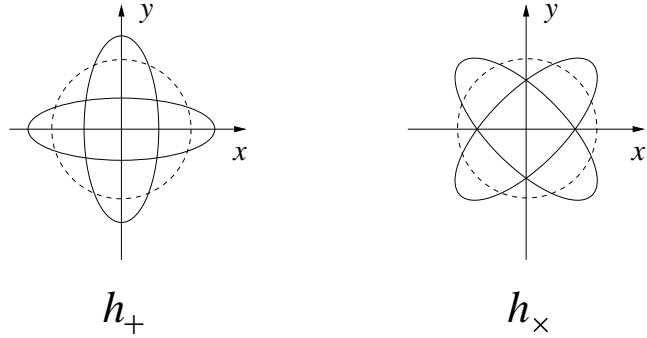}}
\caption{The two gravitational-wave polarisations $h_+$ and $h_\times$.}
\label{poles}\end{figure}
Let us now assume that the waves propagate in the $z$-direction, i.e.
that we have $h_{jk}=h_{jk}(t-z)$. Then one can show that 
we have only two independent components;
\be
h_+ = h_{xx}^{\TT} = -  h_{yy}^{\TT} \qquad
h_\times = h_{xy}^{\TT} =   h_{yx}^{\TT}
\ee
What effect does $h_+$ have on matter? 
Consider a particle initially located at $(x_0,y_0)$ and 
let $h_\times=0$ to find that 
\begin{eqnarray}
\delta x &=& { 1 \over 2} h_+ x_0 \\
\delta y &=& - { 1 \over 2} h_+ y_0 
\end{eqnarray}
That is, if $h_+$ is oscillatory (a wave!) then an object 
will first experience a stretch in the $x$-direction accompanied
by a squeeze in the $y$-direction. One half-cycle later, the 
squeeze is in the $x$-direction and the stretch in the $y$-direction. 
It is straightforward to show that
the effect of  $h_\times$ is  the same, but rotated by 45 degrees.
This is illustrated in Fig.~\ref{poles}.
A general wave will be a linear combination of the two polarisations. 

But wait a second! We are discussing the effect of gravitational \emph{waves}
without actually having proven that Einstein's theory predicts their
existence. To remedy this, consider small perturbations away from a flat
spacetime. That is, use
$g_{\alpha\beta} = \eta_{\alpha\beta} + h_{\alpha\beta}$
to get the (linearised) Ricci tensor:
\be
R_{\mu \nu} = {1 \over 2} ( h_{\alpha \nu , \mu}^{\ \ \ \  \ \alpha} + 
h_{\mu\alpha\ \ \ \nu}^{\ \ \  , \alpha} - h^\alpha_{\ \ \alpha , \mu \nu} 
- h_{\mu \nu , \alpha}^{\ \ \ \ \alpha})
\label{hric}\ee
In vacuum, 
Einstein's equations are equivalent to requiring that 
\be
R_{\mu \nu} = 0
\ee
Now consider what happens if we 
make a coordinate transformation. 
For
\be
x^\alpha  \rightarrow x^\alpha + \xi^\alpha
\ee
we get
\be
h_{\mu \nu} \rightarrow h_{\mu\nu} - \xi_{\mu,\nu}-\xi_{\nu,\mu}
\ee
Use this freedom to impose the harmonic gauge condition 
$h_{\mu \alpha}^{\ \ \ , \alpha} = 0 $. Since the
gauge remains unchanged for any 
transformation such that
$\Box \xi_\mu = 0$, we can impose further conditions. We take 
one of these to be $h_\alpha^{\ \alpha}=0$, to get
\be
 h_{\mu \nu , \alpha}^{\ \ \ \ \ \alpha}  = \Box h_{\mu \nu} = 0
\ee
That is, the metric variations are governed by a standard 
wave equation. Finally, use $h_{00} = h_{j0} = h_{0j} = 0$
to get $ h_{jk} = h_{jk}^{\TT}$ as before. The set of 
coordinates we have introduced is known as TT-gauge.

Before we move on to discuss the modelling of various gravitational-wave sources, 
it is worth elucidating an issue that caused serious debate until
the late 1960s.  We need to demonstrate that gravitational
waves carry energy. This is a tricky problem because, as we have already 
pointed out, one can only deduce the presence of a wave from the
relative effect on two (or more) test particles. This means that one cannot 
localize the wave to individual points in space, and hence cannot 
directly ``measure'' its energy. In order to construct a meaningful
energy expression we need to average over one (or more) wavelengths. 
Defining perturbations with respect to an averaged spacetime metric, i.e.
using
\be
g_{\mu \nu}^{(\B)} = \left< g_{\mu\nu}\right>
\longrightarrow
g_{\mu\nu} = g_{\mu \nu}^{(\B)} + h_{\mu\nu}
\ee
and expanding in powers of $h$ (which is presumed small), we have
the (schematic) Einstein equations, 
\be
\left< G_{\mu \nu} \right>  = G_{\mu\nu}^\B +
\underbrace{\left<G_{\mu\nu}^{(1)}\right>}_{O(h_{\mu \nu})} +  
\underbrace{\left<G_{\mu\nu}^{(2)}\right>}_{ O(h_{..}h_{..})}  = 0
\ee
Here, the term that is linear in $h$ will vanish if we 
average over a wavelength. This means that we can deduce an expression for 
the stress energy-tensor for gravitational
waves:
\be
 G_{\mu\nu}^\B = - 8\pi \left<{ G_{\mu\nu}^{(2)} \over 8\pi}  \right>
\equiv 8\pi T_{\mu \nu}^\GW
\ee
Working out the algebra one can show that, in TT-gauge 
we will have 
\be 
T_{\mu\nu}^\GW = { 1 \over 32\pi} \left< h^\TT_{ij,\mu}  h^{\TT ij}_{\ \ \ \ , \nu} \right>
\ee

In particular, the energy propagating in the $z$-direction then follows from
\be
 T^\GW_{0z} = - T_{00}^\GW = - { 1 \over 16 \pi} \left< \dot{h}_+^2 + \dot{h}_\times^2\right>
\longrightarrow 
\dot{E} = - T^\GW_{0z} 
= { \omega^2 \over 16 \pi} \left< h_+^2 + h_\times^2 \right> 
\ee
where the frequency of the wave is $\omega$.
Finally assuming that $h_+\sim h_\times \sim  h \sin \omega(t-z)$, 
and integrating 
over a sphere with radius $r$, we get
\be
\dot{E} = { \omega^2  r^2 \over 4 } h^2 \rightarrow 
 |\dot{h}|^2 = {  4 G \over  c^3 r^2} \dot{E}
\label{energy}\ee
As we will now demonstrate, this is a very useful relation.

\subsection{Estimating the gravitational-wave amplitude}
\label{sec:gwam}

We can use the expression (\ref{energy}) to infer the gravitational-wave strain 
associated with typical gravitational-wave sources.
Let us characterise a given event by a timescale
$\tau$ and assume that the signal is  monochromatic
(with frequency $f$). Then  
 we can use $dE/dt\approx E/\tau$ and $\dot{h} \approx 2 \pi f h$ to
deduce that
\be
h \approx 5\times 10^{-22} \left( { E \over 10^{-3} M_\odot c^2} \right)^{ 1/2}
\left({ \tau \over 1 \mbox{ ms} }\right)^{-1/2} 
\left( {f\over 1 \mbox{ kHz}} \right)^{-1} \left( { r\over 15 \mbox{ Mpc}}\right)^{-1}
\label{hraw}\ee
If the signal analysis is based on matched filtering, the \emph{effective amplitude} 
improves roughly as the square root of the number of observed cycles $n$. 
Using $n \approx f \tau $ we get 
\be
h_{\rm c} 
\approx 5\times 10^{-22} \left( { E \over 10^{-3} M_\odot c^2} \right)^{ 1/2}
\left({ f \over 1 \mbox{ kHz} }\right)^{-1/2} 
\left( { r\over 15 \mbox{ Mpc}}\right)^{-1}
\label{heff}\ee
This is a crucial expression. We see that 
the ``detector sensitivity'' essentially depends only on the 
radiated energy, the characteristic frequency and the distance to the source.
That is, in order to obtain a rough estimate of the relevance of a 
given gravitational-wave source at a given distance we only need to estimate the
frequency and the radiated energy. Alternatively, if we know the energy released 
can work out the distance at which these sources can be detected.  

It is quite easy to obtain a rough idea of the frequencies involved.
The dynamical frequency of any self-bound system with mass $M$ and radius $R$ 
is
\be
f \approx { 1 \over 2 \pi} \sqrt{ {GM \over R^3} } 
\label{freq}\ee
Thus, the natural 
frequency of a (non-rotating) black hole should be 
\be
f_{\rm BH} \approx 10^4 \left( {M_\odot \over M} \right)\mbox{ Hz}
\ee
Medium sized black holes, with masses in 
the range $10-100M_\odot$, will be prime sources for ground-based 
interferometers, while black holes
with masses $\sim10^6M_\odot$ should radiate in 
the LISA bandwidth. 
Meanwhile, neutron stars, with a mass of $1.4 M_\odot$ 
compressed inside a radius of 
10~km, will radiate at 
\be
f_{\rm NS} \approx 2 \mbox{ kHz}
\ee
This means that one would expect neutron physics to be 
in the range for ground based detectors. In fact, given the likely 
need to detect signals with frequencies above 1~kHz, neutron star 
signals provide a strong motivation for the development of 
(third generation) high-frequency detectors. 

Given that the weak signals are going to be buried in detector noise, 
we need to obtain as accurate theoretical models as possible.
The rough order of magnitude estimates we just derived will certainly 
not be sufficient, even though they provide an indication 
as to whether it is worth spending the time and effort 
required to build a detailed model. Such source models
are typically obtained using either
\begin{itemize}
\item approximate perturbation techniques, eg. expansions in 
small perturbations away from a known solution to the Einstein equations, the 
archetypal case being black-hole and neutron star oscillations.

\item post-Newtonian approximations, essentially an expansion 
in the ratio between a characteristic velocity of the system and the 
speed of light, 
most often used to model
the inspiral phase of a compact binary system.

\item numerical relativity, where the Einstein equations are formulated
as an initial-value problem and solved on the computer. This is the only 
way to make progress in situations where the full nonlinearities of the theory 
must be included, eg. in the merger of black holes and neutron stars
or a supernova core collapse. 

\end{itemize} 

Here we will only describe the first step beyond a Newtonian 
description, where the gravitational radiation is described 
by the so-called \emph{quadrupole formula}. 
For a source with weak internal gravity we have (in TT-gauge)
\be 
\Box h_{\mu \nu} = - 16 \pi T_{\mu \nu}
\ee
We can solve this equation using
the standard retarded Green's function to get
\be
h_{\mu \nu} (t, \vec{x}) =  4 \int { T_{\mu \nu} 
( \vec{x}^\prime, t^\prime = t - |\vec{x}-\vec{x}^\prime|) 
 \over |\vec{x}-\vec{x}^\prime| } d^3 x
\ee
Matching of the near-zone solution to an outgoing wave solution far away
leads to the expression 
\be
h_{jk}^\TT =  { 2 G \over r c^4 }  \ddot{\Ibar}^\TT_{jk}(t-r)
\ee
where 
\be
\Ibar_{jk} \equiv \int \rho \left( x_j x_k - { 1\over 3} r^2 \delta_{jk} \right)   d^3 x
\ee
is the reduced quadrupole moment of the source. 
Consider a system of mass $M$ with typical internal velocity $v$. 
Then we see that 
\be
h \approx \left( {2GM \over c^2 }\right) \left( { v\over c}\right)^2 { 1 \over r}  
\ee
which shows (no surprise!)
that in order to generate strong gravitational waves  
we need large masses moving at high speeds.

From the formulas we derived earlier, we find that energy is radiated at 
a rate
\be
{ dE \over dt}
 = 
- { G \over 5 c^5 } \left< \dddot{\Ibar}_{jk} \dddot{\Ibar}^{jk} \right>
\ee
The radiated angular momentum follows from the (usually) 
weaker current multipole radiation, which is 
governed by a similar expression.
\begin{figure}
 \centerline{\includegraphics[width=8.5cm]{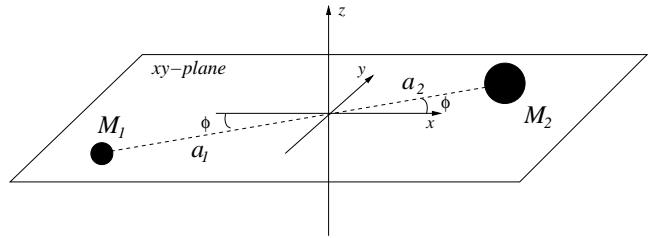}}
\caption{A schematic illustration of a compact binary system.}
\label{binfig}\end{figure}
Let us now apply the above results to the 
potentially most important gravitational-wave source,  a compact binary system.
Gravitational waves are emitted as the stars (or black holes) 
orbit each other and as a result the binary separation decreases. 
Consider a binary system with individual masses $M_1$ and $M_2$ and 
separation $2 R$.  Introduce the total and reduced masses
\be
M = M_1 + M_2 \qquad \mbox{ and } \qquad \mu = { M_1 M_2 \over M}
\longrightarrow
M_1 a_1 = M_2 a_2 = \mu R
\ee
and work in the coordinate system illustrated in figure~\ref{binfig}.
Working out the required (time-varying)
components of the quadrupole moment, we have
\be
\Ibar_{xx} = -\Ibar_{yy} = { \mu R^2 \over 2}  \cos 2 \phi  \quad
\mbox{ and } \quad
\Ibar_{xy} = \Ibar_{yx} = { \mu R^2 \over 2}  \sin 2 \phi
\ee
and we find that 
\be
{ dE \over dt} = { G \over 5 c^5 } \left< \dddot{\Ibar}_{jk} \dddot{\Ibar}^{jk} \right> =  { 32 \over 5} { G \over c^5} \mu^2 R^4 \Omega^6
\ee
Next, 
determining the orbital rotation frequency 
from Kepler's law $\Omega^2 = { M/ R^3}$,  
and introducing the so-called ``chirp mass''
${\cal M} = \mu^{3/5} M^{2/5}$
we have the final result
\be 
{ dE \over dt} = { 32 \over 5} { G^4 \over c^5}   ( {\cal M} \Omega )^{10/3}
\label{eloss}\ee
\begin{figure}
\vskip 0.4cm
 \centerline{\includegraphics[height=6.5cm]{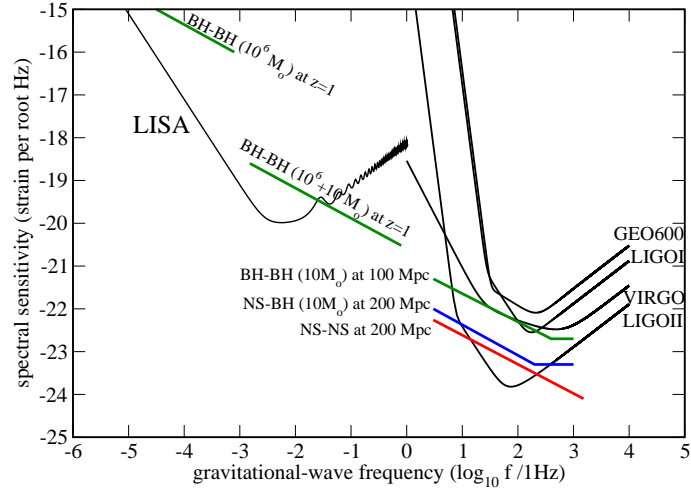}}
\medskip
\caption{Estimated signal strengths for various inspiralling binaries relevant for 
ground- and space-based detectors. } 
\label{bins}\end{figure}
Moreover, from
\be
h_{xx}^\TT \approx
- { 2 {\cal M}^{5/3} \Omega^{2/3} \over r} \cos 2\Omega (t-r)
\label{h_ampl}
\ee
we can estimate the effective amplitude of the binary signal:
\be
h_c \approx \sqrt{ f t }\  h \sim { M \over r} \left( {R\over M}\right)^{1/4} 
\ee
This shows that, even though the 
actual signal gets stronger, its detectability decreases as the orbit shrinks.
Figure~\ref{bins} compares our estimated gravitational-wave strain to the predicted
noise-curves for various gravitational-wave detectors. Essentially, one would expect that
\begin{itemize}
\item Advanced LIGO may observe several binary systems per year. 

\item The space-based LISA detector 
may suffer an ``embarrassment of riches'', with a large number of known galactic
binaries leading to detectable signals (most likely generating 
to a ``binary noise'' which will be 
difficult to filter out). 
\end{itemize}

To predict the rate at which the binary orbit shrinks as a result
of gravitational-wave emission
we need to estimate to total energy of the system:
\be
E = { M_1 v_1^2 \over 2} + { M_2 v_2^2 \over 2} - { G M_1 M_2 \over R}  
= - { G \mu M \over 2R}  = - G { {\cal M}^{5/3} \Omega^{2/3} \over 2}
\ee
From this we find that the period  $P$ of the system changes as
\be
{ \dot{P} \over P} = - {96 \over 5} { G^3 \over c^5} {\cal M}^{5/3} \Omega^{8/3}
\ee
For the {\em binary pulsar} 1913+16, the predicted change in orbital 
period agrees with the theoretical prediction to within 1\%. 
This indirect proof that gravitational waves exist led to Hulse and Taylor
being awarded the 1993 Nobel prize in physics.

Finally, note that that the  chirp-mass $\cal M$ plays
 a key role in our various expressions. 
From 
\be
{ \dot{P} \over P} \sim {\cal M}^{5/3} \Omega^{8/3} \quad 
\mbox{ and } \quad h \sim  {  {\cal M}^{5/3} \Omega^{2/3} \over r} 
\ee
we see that this is the only combination of the 
two masses that can be infered from the signal (at this level of approximation: 
higher order post-Newtonian corrections depend on the individual
masses, the spins etcetera). 
Suppose that we detect both the change in the amplitude $h$ and the shift in the 
frequency. Then one can infer both the chirp mass and  the 
distance to the source, and in effect 
coalescing binaries are {\em standard candles} which 
can be used to constrain cosmological parameters.

One often classifies
gravitational-wave sources by the nature of the waves. This is convenient 
because the different classes require different approaches to the data-analysis
problem;

\begin{itemize}

\item {\em Chirps}. As a binary system radiates gravitational waves and 
loses energy the two constituents spiral closer together. As the separation
decreases the gravitational-wave amplitude increases, leading to a 
characteristic ``chirp'' signal. 

\item {\em Bursts}. Many scenarios lead to burst-like gravitational waves.
A typical example would be black-hole oscillations excited during binary 
merger. 

\item {\em Periodic}. Systems where the gravitational-wave backreaction
leads to a slow evolution (compared to the observation time)
may radiate persistent waves with a virtually constant frequency.
This would be the gravitational-wave analogue of the radio 
pulsars. 

\item {\em Stochastic}. A stochastic (non-thermal) 
background of gravitational waves
is expected to have been generated following the Big Bang. 
 One may also have to deal with 
stochastic gravitational-wave signals when the sources are too 
abundant for us to distinguish them as individuals.

\end{itemize}

\section{High-frequency gravitational wave sources}

Having introduced the key concepts required for a discussion of 
gravitational-wave physics we will now focus our attention on 
sources that radiate above  a hundred Hz or so, i.e. which may 
at least in principle be detectable from the ground. As we will see
there are strong motivations for constructing detectors which are 
sensitive up to  (ideally) several kHz.  

\subsection{Radiation from binary systems}

It is easy to understand why 
binary  systems  are considered the ``best'' sources of  gravitational
waves. They  emit  copious  amounts   or gravitational
radiation,  and  for  a given  system we know quite accurately
the amplitude and  frequency of the gravitational waves in
terms of the masses of the two bodies and their separation (see
section \ref{sec:gwam}). 

The gravitational-wave signal from inspiraling binaries is 
approximatelly sinusoidal, see equation (\ref{h_ampl}), with a frequency 
which is twice the orbital frequency of the binary. As the binary system 
evolves the orbit shrinks and the frequency increases in the characteristic chirp. 
Eventually, depending on the masses of the binaries, the frequency of the 
emitted gravitational waves will enter the bandwidth of the detector 
at the  low-frequency end and will evolve quite fast towards higher frequencies. 
A system consisting  of  two  neutron  stars
will  be detectable by LIGO when the frequency of the
gravitational waves is  $\sim$10Hz until the final coalescence
around 1 kHz. This process will last for about 15 min and the
total number of observed cycles will be of the order of $10^4$,
which leads to  an enhancement of the detectability by  a  factor
100 (remember $h_c \sim \sqrt{n}h$). 
Binary neutron  star systems and  binary  black  hole systems
with masses of the order of 50M$_\odot$ are  the primary sources
for LIGO. Given the anticipated sensitivity of LIGO, binary black
hole systems are the most  promising sources and could be detected
as far as 200 Mpc  away.  For the present
estimated sensitivity of LIGO the event rate 
is probably a few per year,
but future improvements of detector sensitivity (the LIGO II
phase) could lead  to  the detection  of  at  least one event per
month.  Supermassive black hole  systems of a few million solar
masses  are  the primary source for LISA. These binary systems are
rare,  but due  to  the huge amount of energy released, they
should  be detectable  from as far away as the boundaries of the
observable universe. Finally, the recent discovery of the highly relativistic 
binary pulsar J0737-3039 \cite{Burgay03} enchanced considerably 
the expected coalescence event rate of NS-NS binaries \cite{Kalogera04}. 
The event rate for initial LIGO is in the best case 
0.2 per year while advanced LIGO might be able to detect 20-1000 events per year.

\subsection{Gravitational collapse}

One of the most spectacular events in the Universe is the
supernova (SN) collapse to create a neutron star (NS) or a black
hole (BH). Core collapse is a very complicated event and a proper
study demands a deep understanding of neutrino
emission, amplification of the magnetic fields,  angular momentum
distribution, pulsar kicks, etc. There are many viable
models for each of the above issues but it is still not
possible to combine all of them together into a consistent explanation.
Gravitational waves emanating from the very first moments of the
core collapse might shed light on all the above problems and help
us understand the details of this dramatic event. Gravitational
collapse compresses matter to nuclear densities, and is
responsible for the core bounce and the shock generation. The
event proceeds extremely fast, lasting less than a second,
and the dense fluid undergoes motions with relativistic speeds
($v/c\sim 0.2-0.4$). Even small deviations from spherical symmetry
during this phase can generate significant amounts of gravitational waves. However,
the size of these asymmetries is not known.
 From observations in the
electromagnetic spectrum we know that stars more massive than
$\sim 8M_\odot$ end their evolution in core collapse and that $\sim 90\%$ of them
are stars with masses $\sim 8-20M_\odot$. During the collapse most
of the material is ejected and if the progenitor star has a mass
$M\lesssim 20M_\odot$ it leaves behind a neutron star. If
$M\gtrsim 20M_\odot$ more than 10\% 
falls back and pushes the
proto-neutron-star (PNS) above the maximum NS mass leading to the
formation of a black hole ({\em type II collapsars}). Finally, if
the progenitor star has a mass $M\gtrsim 40M_\odot$ no supernova
takes place. Instead, the star collapses directly to  a BH ({\em type I
collapsars}).

A significant amount of the ejected material can fall back,
subsequently spinning up and reheating the nascent NS.
Instabilities can be excited during such a process. If a BH
was formed, its quasi-normal modes (QNM) can be excited for as long as
the process lasts. ``Collapsars'' accrete material during the very
first few seconds, at rates $\sim 1-2M_\odot$/sec. Later the
accretion rate is reduced by an order of magnitude but still
material is accreted for a few tenths of seconds. Typical
frequencies of the emitted gravitational waves are in the range 1-3kHz for $\sim
3-10 M_\odot$ BHs. If the disk around the central object has a
mass $\sim 1M_\odot$ self-gravity becomes important and
gravitational instabilities (spiral arms, bars) might develop and
radiate gravitational waves. 
Toroidal configurations can be also formed around the collapsed 
object. Their instabilities and oscillations might be an interesting 
source of gravitational waves\cite{Zannoti}.
There is also the possibility that the collapsed
material might fragment into clumps, which orbit for some cycles
like a binary system ({\em fragmentation instability}~\cite{Fryer2001}).

The supernova event rate is 1-2 per century per galaxy
\cite{Cappellaro} and about 5-40\% 
of them produce BHs through the
fallback material \cite{FryerKalogera}. Conservation of angular
momentum suggests that the final objects should rotate close to
the mass shedding limit, but this is still an open question, since
there is limited knowledge of the initial rotation rate of the
final compact object. Pulsar statistics suggest that the initial
periods are probably considerably shorter than $20$~ms.  This
strong increase of rotation during the collapse has been observed
in many numerical simulations (see e.g. \cite{FryerHeger,
DFM2002b}).

Core collapse as a potential source of gravitational waves has been studied for
more than three decades (some of the most recent calculations can
be found in \cite{Finn90, Zwerger1997, Rampp1998, Fryer2002, DFM2002b,
Ott2003, Kotake2003, Shibata03, Whisky}). All these numerical calculations show that
signals from Galactic supernova ($d\sim 10$kpc) are detectable
even with the initial LIGO/EGO sensitivity at frequencies
$\lesssim$1kHz. Advanced interferometers can detect signals from distances of
1~Mpc but it will be difficult with the designed broadband
sensitivity to resolve signals from the Virgo cluster
($\sim$15Mpc). The typical gravitational wave amplitude from the 2D numerical
simulations \cite{DFM2002b, Ott2003} for an observer located on
the equatorial plane of the source is
\begin{equation}
h\approx 9 \times 10^{-21}\varepsilon \left({ 10 {\rm kpc} \over
d}\right)
\end{equation}
where $\varepsilon \sim 1$ is the normalized gravitational wave amplitude. The
total energy radiated in gravitational waves during the collapse is $\lesssim
10^{-6}-10^{-8} M_\odot c^2$. However, these numerical estimates are not
yet conclusive, as important aspects such as 3D hydrodynamics
combined with proper spacetime evolution have been neglected. The
influence of the magnetic fields have been ignored in most
calculations. The proper treatment of these issues might not
change the above estimates by orders of magnitude but it will
provide a conclusive answer. There are also issues that need to be
understood such as the pulsar kicks (velocities even higher than
1000 km/s) which suggest that in a fraction of newly-born NSs (and
BHs) the process may be strongly asymmetric 
\cite{Caraveo93, Burrows1996, Muller97, Spruit98,Loveridge}. 
Also, the polarization
of the light spectra in supernovae indicates significant
asymmetries \cite{Wheeler1999}. Better treatment of the
microphysics and construction of accurate progenitor models for
the angular momentum distributions are needed. All these issues
are under investigation by many groups.

{\em Accretion Incuded Collapse} (AIC) is also a possible source of high 
frequency gravitational waves. AIC takes place when a white dwarf (WD) 
exceeds the Chandrasekhar limit due to accretion of material and begins to collapse. 
The cooling via neutrino emission 
does not reduce the heating significantly and the 
collapsing WDs reach appropriate temperatures for ignition of nuclear burning (Type Ia supernova). 
Estimates suggest that about $0.1M_\odot$ material is ejected.
Since the WD is pushed over the Chandrasekhar limit due to accretion, 
it will rotate fast enough to allow various types of instabilities \cite{Lindblom01}. 
The galactic rate of accretion induced collapse 
is about $10^{-5}$/yr which means that AIC are about 
1000 times rarer than core collapse SN.

\subsection{Rotational instabilities}

Newly born neutron stars are expected to rotate rapidly enough to be
subject to rotation induced instabilities. These instabilities arise from
non-axisymmetric perturbations having angular dependence $e^{i
m\varphi}$. Early Newtonian estimates have shown that a {\em
dynamical bar-mode ($m=2$) instability} is excited if the ratio
$\beta=T/W$ of the rotational kinetic energy $T$ to the
gravitational binding energy $W$ is larger than $\beta_{\rm dyn}=
0.27$. The instability develops on a dynamical time scale (the
time that a sound wave needs to travel across the star) which is
about one rotation period, and may last from 1 to 100 rotations
depending on the degree of differential rotation in the PNS.
Another class of instabilities are those driven by dissipative
effects such as fluid viscosity or gravitational radiation. Their
growth time is much longer (many rotational periods) but they can
be excited for significantly lower rotational rates, $\beta
\gtrsim 0.14$ in the case of the fundamental modes of oscillation
of the star.

\subsection{Bar-mode instability}

The dynamical bar-mode instability can be excited in a hot PNS, a few
milliseconds after the core-bounce, given a sufficiently large
$\beta$. It might also be excited a few tenths of seconds later,
when the NS cools enough due to neutrino emission and contracts
still further ($\beta \sim 1/R$). The amplitude of the emitted
gravitational waves can be estimated as $h\sim M R^2 \Omega^2/d$,
where $M$ is the mass of the body, $R$ its size, $\Omega$ the
rotational rate and $d$ the distance from Earth. This leads to an
estimate of the gravitational wave amplitude
\begin{equation}
h\approx 9\times 10^{-23} \left({\epsilon \over 0.2} \right)
\left({f\over 3 {\rm kHz}}\right)^2 \left({15 {\rm Mpc} \over
d}\right) M_{1.4} R_{10}^2.
\end{equation}
where $\epsilon$ measures the ellipticity of the bar. Note that
the gravitational wave frequency $f$ is twice the rotational 
frequency $\Omega$.
Such a signal is detectable only from sources in our galaxy or the
nearby ones (our Local Group). If the sensitivity of the detectors
is improved in the kHz region, then signals from the Virgo cluster
may be detectable. If the bar persists for many ($\sim$ 10-100)
rotation periods, then even signals from distances considerably
larger than the Virgo cluster could be detectable, 
cf. Figure~\ref{barmode}. 
The event rate is of the same order as the SN rate 
(a few events per century per galaxy): this means that given 
the appropriate sensitivity at frequencies between 1-3kHz we 
might be able to observe a few events per year. 
The bar-mode instability may also be excited during the merger of
NS-NS, BH-NS, BH-WD and even in type II collapsars (see discussion in
\cite{Kobayasi2003}).

In general, the above estimates rely on Newtonian hydrodynamics
calculations; GR enhances the onset of the instability slightly,
$\beta_{\rm dyn}\sim 0.24$ \cite{SBS2000} and $\beta_{\rm dyn}$
may be even lower for large values of the compactness (larger
$M/R$). The bar-mode instability may be excited for significantly
smaller $\beta$ if centrifugal forces produce a peak in the
density off the sources rotational center\cite{Centrella2001}.
Rotating stars with a high degree of differential rotation are
also dynamically unstable for significantly lower $\beta_{\rm
dyn}\gtrsim 0.01$ \cite{Shibata2002, Shibata2003}. 
In this scenario
the unstable neutron star settles down to a non-axisymmetric
quasistationary state which is a strong emitter of quasi-periodic
gravitational waves
\begin{equation}
h_{\rm eff} \approx 3\times 10^{-22} \left({R_{\rm eq} \over 30
{\rm km}} \right) \left({f\over 800 {\rm Hz}}\right)^{1/2}
\left({100 {\rm Mpc} \over d}\right) M_{1.4}^{1/2} .
\end{equation}
The bar-mode instability of differentially rotating neutron stars
could be an excellent source of gravitational waves provided that
the dissipation of non-axisymmetric
perturbations by viscosity and magnetic fields is negligible.
That this is the case is far from clear.
Magnetic fields might actually enforce the uniform rotation of the
star on a dynamical timescale and the persistent non-axisymmetric
structure might not have time to develop at all.

Numerical simulations have shown that the $m=1$ one-armed spiral
mode might become dynamically unstable for considerably lower
rotational rates \cite{Centrella2001, SBM2003}. This $m=1$
instability depends critically on the softness of the equation of
state (EoS) and the degree of differential rotation.

\subsection{CFS instability, f- and r-modes}

After the initial bounce, neutron stars may maintain a
considerable amount of deformation. They settle down to an
axisymmetic configuration mainly due to emission of gravitational waves, viscosity
and magnetic fields. During this phase QNMs are excited.
Technically speaking, an oscillating non-rotating star has equal
values $\pm |\sigma|$ (the frequency of a mode) for the forward
and backward propagating modes (corresponding to $m=\pm|m|$).
Rotation changes the mode frequency by an amount $\delta \sigma
\sim m \Omega$ and both the prograde and retrograde modes will be
dragged forward by the stellar rotation. If the star spins
sufficiently fast, the originally retrograde 
mode will appear to be moving forwards in the
inertial frame (according to an observer at infinity), but still backwards in
the rotating frame (for an observer rotating with the star). Thus, an
inertial observer sees gravitational waves with positive angular momentum emitted
by the retrograde mode, but since the perturbed fluid rotates
slower than it would in absence of the perturbation, the angular
momentum of the mode itself is negative. The emission of gravitational waves
consequently makes the angular momentum of the mode increasingly
negative  leading to an instability. From the above, one can
easily conclude that a mode will be unstable if it is retrograde
in the rotating frame and prograde for a distant observer
measuring a mode frequency $\sigma-m\Omega$ i.e. the criterion
will be $ \sigma(\sigma-m\Omega) < 0$.

This class of {\em frame-dragging instabilities} is usually
referred to as Chandra-sekhar--Friedman--Schutz (CFS) 
\cite{Chandra70, FS78} instabilities.
For the high frequency ($f$ and $p$) modes the instability 
is possible only
for large values of $\Omega$ or for quite large $m$. In general,
for every mode there will always be a specific value of $\Omega$ for
which the mode will become unstable, although only modes with
$|m|<5$ have an astrophysically significant growth time. The CFS
mechanism is not only active for fluid modes but also for the {\em
spacetime} or the so-called {\em w}-modes\cite{KRA2002}. It is
easy to see that the CFS mechanism is not unique to gravitational
radiation: any radiative mechanism will have the same effect.

In GR, the {\em f-mode} ($l=m=2$) becomes unstable for $\beta
\approx 0.06-0.08$ \cite{SF1998}. If the star has significant
differential rotation the instability is excited for somewhat
higher values of $\beta$ (see e.g. \cite{Yoshida,
Stergioulas2003}). The $f$-mode instability is an excellent source
of gravitational waves. After the brief dynamical phase, the PNS becomes unstable
and the instability deforms the star into a non-axisymmetric
configuration via the $l=2$ bar mode. Since the star loses angular
momentum it spins down, and the gravitational wave frequency sweeps from 1~kHz
down to about 100~Hz \cite{LaiShapiro95}. If properly modelled 
such a signal can be
detected from a distance of 100~Mpc (if the mode grows to a large
nonlinear amplitude).

\begin{figure}
\vskip 0.7cm
\centerline{\includegraphics[height=6.5cm,width=7.5cm]{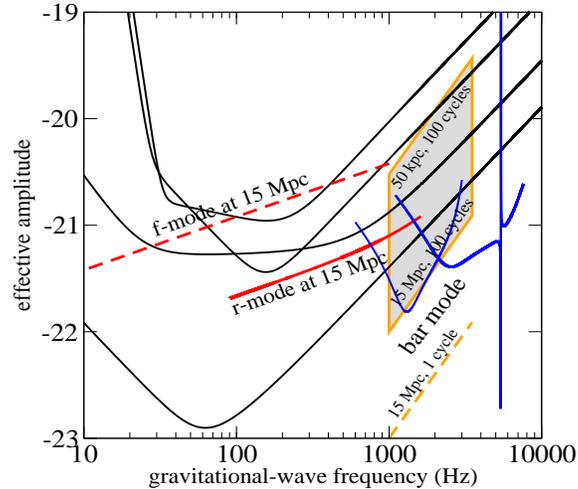}}
\medskip
\caption{The estimated strength of gravitational waves from the dynamical 
bar-mode instability and the CFS instability of the f- and r-modes. 
The estimates are compared to the predicted noise of the various 
interferometers and also the possible noise curve for a dual 
cylinder detector. }
\label{barmode}
\end{figure}

Rotation  not only shifts the frequencies of the various modes; it also
gives rise to  the Coriolis force,
and an associated new family of {\em rotational } or {\em
inertial} modes. Inertial modes are primarily velocity
perturbations. Of special interest is the quadrupole inertial
mode ($r$-mode) with $l=m=2$. The frequency of the $r$-mode in the
rotating frame of reference is $\sigma = 2 \Omega /3$. Using the
CFS criterion for stability we can easily show that the $r$-mode
is unstable for any rotation rate of the star. For temperatures
between $10^{7}-10^{9}$K and rotation rates larger than 5-10\% of
the Kepler limit, the growth time of the unstable mode is much shorter
than the damping times due to  bulk and shear viscosity. The mode
grows until it saturates due to non-linear effects \cite{LOM98,Owen98,AKS99}. 
The strength of the emitted gravitational waves depends on the saturation 
amplitude $\alpha$. Mode coupling might not
allow the growth of the instability to  amplitudes larger than $\alpha
\approx 10^{-2}-10^{-3}$ \cite{Arras2002}. The existence of a
crust \cite{Bildsten99, LOU2000} or of hyperons in the core \cite{LO2002}  and strong
magnetic fields \cite{Rezzolla2000}, affect the efficiency of the 
instability (for extended reviews see \cite{AK2001, Nils2003}). 
For newly-born neutron stars
the amplitude of gravitational waves might not be  such that the signals
will be detectable only from the local group of galaxies ($d<1$Mpc)
\begin{equation}
h(t)\approx 10^{-21} \alpha \left( {\Omega \over 1 {\rm
kHz}}\right)\left({100 {\rm kpc} \over d}\right)
\end{equation}
see Figure~\ref{barmode}.

If the compact object is a strange star, then the r-mode instability will
not reach high amplitudes ($\alpha \sim 10^{-3}-10^{-4}$) but it
will persist for a few hundred years and in this case there might
be up to ten unstable stars per galaxy radiating gravitational waves
at any time \cite{AJK2002}.
Integrating data for a few weeks can then lead to an effective
amplitude $h_{\rm eff}\sim 10^{-21}$ for galactic signals at
frequencies $\sim 700-1000$Hz. The frequency of the signal changes
only slightly on a timescale of a few months, so the radiation is
practically monochromatic.

Old accreting neutron stars, radiating gravitational waves due to the $r$-mode
instability, at frequencies 400-700Hz, are probably a better
source \cite{AKS1999, AJKS2000, Heyl, Wagoner}. Still, the
efficiency and the actual duration of the process depends on the
saturation amplitude $\alpha$. If the accreting compact object is
a strange star or has a hyperon core then it might be a persistent source which radiates
gravitational waves for as long as accretion lasts~\cite{AJK2002, Reisenegger}.

\subsection{Oscillations of black holes and neutron stars}

{\em Black-hole ringing}. The merger of two neutron stars or black holes  
or the collapse of a supermassive star (collapsar of type I or II) will produces a black hole.
The newly formed black hole will ring, emitting a characteristic signal until it 
settles down to the stationary Kerr state. This characteristic signal, the so-called 
quasi-normal mode oscillation, will be a unique probe of the black hole's  existence.
Although the ringing phase does not last very long (a few tenths of a ms), 
the ringing due to the excitation by the fallback material 
might last for secs\cite{Araya03,Nagar04}. The frequency and damping time of the 
black-hole ringing for 
the $l=m=2$ oscillation mode can be estimated via the relations \cite{Echeverria}
\begin{eqnarray}
\sigma&\approx& 3.2 {\rm kHz} \
M_{10}^{-1}\left[1-0.63(1-a/M)^{3/10}\right] \\
 Q&=&\pi \sigma
\tau \approx 2\left(1-a\right)^{-9/20} \label{bhqnm}
\end{eqnarray}
These relations together with similar ones either for the 2nd QNM
or the $l=2$, $m=0, \pm 1$ can uniquely determine the mass $M$ and
angular momentum $a$ of the BH if the frequency and the damping
time of the signal have been accurately extracted \cite{Finn,
Dryer}. The amplitude of the ring-down waves depends on the BH's
initial distortion. If the excitation of the BH is due to infalling
material then the energy is roughly $\Delta E \gtrsim \epsilon \mu
c^2(\mu/M)$ where $\epsilon \gtrsim 0.01$ \cite{Davis1971}. This
leads to an effective gravitational wave amplitude
\begin{equation}
h_{\rm eff}\approx 2\times 10^{-21}\left({\epsilon \over 0.01}
\right)\left({10 {\rm Mpc}\over d}\right)\left( {\mu\over
M_\odot}\right)
\end{equation}
This approximate result has been verified by more detailed full non-linear 
simulations \cite{Anninos}

\medskip
{\em Neutron star ringing}. If the collapse leaves behind a
compact star, various types of oscillation modes might be excited
which can help us estimate parameters of the star such as radius,
mass, rotation rate and EoS\cite{AK1998, KAA2001, AC2001}. This
{\em gravitational wave asteroseismology} is a unique way to find
the radius and the EoS of compact stars. One can derive
approximate formulas in order to connect the observable
frequencies and damping times of the various stellar modes to the
stellar parameters. For example, for the fundamental oscillation
($l=2$) mode ($f$-mode) of non-rotating stars we get \cite{AK1998}
\begin{eqnarray}
\sigma({\rm kHz})&\approx& 0.8+1.6 M_{1.4}^{1/2}R_{10}^{-3/2}
+ \delta_1 m{\bar \Omega} \\
\tau^{-1}({\rm secs}^{-1})&\approx&
M_{1.4}^3R_{10}^{-4}\left(22.9-14.7M_{1.4}R_{10}^{-1}\right)+
\delta_2 m {\bar \Omega}
\end{eqnarray}
where ${\bar \Omega}$ is the normalized rotation frequency of the
star, and $\delta_1$ and $\delta_2$ are constants estimated by
sampling data from various EoS. The typical frequencies of the NS
modes are higher than 1~kHz. On the other hand, 2D simulations of
rotating core-collapse have shown that if a rapidly rotating NS is
created, then the dominant mode is the quasi-radial mode
(``$l=0$''), radiating through its $l=2$ piece at frequencies
$\sim$  800Hz-1kHz \cite{DFM2002b}. Since each type of mode is
sensitive to the physical conditions where the amplitude of the
mode eigenfunction is greatest,  the more information we get from the various
classes of modes the better we will understand the details of the star.

Concluding, we should mention that the tidal disruption of a NS by
a BH \cite{Vallisneri} or the merging of two NSs \cite{Rasio} may
give valuable information for the radius and the EoS if we can
detect the signal at frequencies higher than 1 kHz.

\medskip
This work has been supported by the EU Programme 'Improving the
Human Research Potential and the Socio-Economic Knowledge Base'
(Research Training Network Contract HPRN-CT-2000-00137). 
KK acknowledges support throught the Center of Gravitational Wave Physics, 
which is funded by the NSF number cooperative agreement PHY 01-14375.


\end{document}